# Human Activity and Mobility Data Reveal Disparities in Exposure Risk Reduction Indicators among Socially Vulnerable Populations during COVID-19

Natalie Coleman[1], Xinyu Gao[1], Jared DeLeon[2], Ali Mostafavi[1]

[1] Zachry Department of Civil and Environmental Engineering, Texas A&M University.
[2] Department of Computer Science & Engineering, Texas A&M University.
Email addresses: ncoleman@tamu.edu (N. Coleman), xy.gao@tamu.edu (X. Gao), jareddeleon@tamu.edu (J. DeLeon), amostafavi@civil.tamu.edu (A. Mostafavi).

## ABSTRACT

Non-pharmacologic interventions (NPIs) are one method to mitigate the spread and effects of the COVID-19 pandemic in the United States. NPIs, which promote protective actions to reduce exposure risk including stay-at-home policies and essential service mandates, can reduce and change mobility patterns within communities. The growing research literature suggests that socially vulnerable populations are disproportionately impacted with higher infection and higher fatality rates associated with the pandemic, though there is limited understanding of the underlying mechanisms to this health disparity. Thus, this research examines two distinct and complimentary datasets at a granular scale through statistical and spatial analyses to extensively understand the exposure risk reduction of various socially vulnerable populations due to NPIs. Our analysis includes five urban areas during the first wave of the COVID-19 pandemic (from January 1, 2020, to July 31, 2020). The ***mobility dataset*** tracks population movement within and between ZIP codes; it is used for an origin-destination network analysis. The ***population activity dataset*** is based on the number of visits from census block groups (CBG) to points of interest (POIs), such as grocery stores, restaurants, education centers, and medical facilities; it is used for network analysis of population-facilities interactions. The ***mobility dataset*** showed that, at the beginning of the year, community members had similar patterns of movement; however, after the implementation of NPIs, socially vulnerable populations engaged in increased mobility in the form of inflow from home ZIP code areas to other ZIP code areas. Similarly, ***population activity*** analysis showed an increased exposure risk for socially vulnerable populations based on a greater number of inflow visits of CBGs to POIs, which increases the risk of contact at POIs, and a greater number of outflow visits from POIs to home CBGs, which increases risk of transmission within CBGs The findings pinpoint variations in exposure risk reduction indicators implied by NPIs among low-income and racial and ethnic minority populations. These findings can assist emergency planners and public health officials in comprehending how different groups are able to implement protective actions associated with NPIs to reduce their exposure risk. The findings can inform more equitable and data-driven NPI policies for future epidemics.

# Introduction

Since the first COVID-19 case in the United States was reported on January 21, 2020 in Snohomish County, Washington[1], the SARS-CoV-2 virus has rapidly spread across the country. As of June 1, 2021, COVID-19 more than 33 million confirmed cases have been diagnosed and approximately 591,000 deaths are attributed to the disease in the United States. To decrease the contact and transmission rate of COVID-19, many states implemented state- or local-level stay-at-home policies as well as the closure of non-essential services starting mid- March 2020. This is because non-pharmacologic interventions (NPIs), which encourage protective actions via social distancing and sheltering-in-place, are effective measures to slow down the spread of COVID-19 [2-4].

Life and daily movement patterns were altered by this pandemic. According to guidelines associated with NPIs, places regarded as non-essential such as schools, gyms, bars, and other commercial complexes, were temporally closed, and mass gatherings and celebratory events were cancelled or postponed. People also tried to curtail their daily essential activities (e.g., refueling cars, purchasing goods) to decrease the risk of infection. Such reduction in movement could be a proxy measurement for the protective actions for reducing exposure risk. To understand the influence and effectiveness of such social distancing practices, many studies have analyzed real-time movement data at multiple scales: country level [5, 6], county level [7-10], and the city level [11, 12]. These studies show that the implementation of non-pharmacologic interventions significantly reduced human activities, and thus reducing possible transmission of virus. The effect of COVID-19 and the NPIs, however, can vary among different subpopulations. Studies of such highly-aggregated human movement and mobility data may have missed the critical disparity among different socio-demographic groups[13].

Socio-demographic disparity has historically been related to societal issues such as disaster recovery, educational resources, and health inequalities [14-19]. In the ever-evolving research literature of the COVID-19 pandemic, earlier studies and reports have captured the disparate impacts associated with different socio-demographic groups at both the community and individual level. For example, Benitez and Yelowitz [20] found that predominately Black and Hispanic neighborhoods had higher COVID-19 cases per capita and higher observed fatalities. Similarly, Abedi et. al[21] concludes that counties with more diverse demographics, such as those with larger population, larger percentage of minority households, lower educational attainment, lower income, or higher disability rates are at a higher risk of COVID-19 infection, and in particular, African Americans are more vulnerable to COVID-19 than other ethnic groups. At the county level, Ossimetha [22] found that counties with socioeconomic disadvantages and less reduced mobility had greater growth in COVID-19 cases and deaths along. Borgonovi and Andrieu [23] found that counties whose residents present with pre-existing medical conditions and low levels of community social capital were more susceptible to experiencing increased rate of infection of COVID-19 due to a more modest in mobility, even suggesting that social distancing practices were related to behavioral changes in mobility. These studies emphasized the exposure and inherent risk disparity of vulnerable subpopulations; however, only a limited number of studies have investigated the extent to which exposure risk reduction conferred by NPIs varied across these sub-populations. Evaluating the exposure risk reduction indicators (based on granular human mobility and activity datasets) conferred by NPIs and their variation across subpopulations may hold the key to understanding the exposure disparities among low-income and racial and ethnic minority groups.

Part of the limitation in studying the effects of NPIs is that current published research focuses on human movement and COVID-19 outcomes at highly aggregated levels (i.e., state- or county level). High-level disparity analysis may ignore an important part of the variation as residential segregation by sociodemographic characteristics can be significant in finer spatial scales. Studies of social vulnerability warn that coarse-scale analysis may fail to detect critical instances of disparities, such as those prominent in inner cities [24]. In fact, finer-scale analysis may yield different results compared to coarser-scale analysis as observed by the law of averages [13, 25]. Studies focusing on fine-scale analysis of disparities in movements and activity reduction of different subpopulations in the context of COVID-19 are rather limited. Among these studies, Benitez and Yelowitz [20] conducted racial and ethnic disparity analysis in COVID-19 cases per capita at the ZIP-code level for six cities, and the findings support that Black and Hispanic populations are correlated with higher rates of COVID-19 cases. The study acknowledges a knowledge gap related to the underlying mechanisms leading to such risk disparities and emphasizes a need to understand such disparities at a granular level. In addition, even among the limited existing studies, the majority have analyzed single mobility and/or population activity datasets. This limits the ability to holistically understand different indicators of exposure risk to the COVID-19 virus. Since each dataset might have limitations in terms of aspects of mobility movements and population activities captured, it is essential to conduct studies with different datasets with intentionality to dissect, interpret, and integrate multiple indicators of exposure risk.

Thus, this research study addresses the knowledge gap by examining the disparities associated with the protective actions to reduce the risk of transmission and by using the mobility and population activity patterns of socially vulnerable groups. Through network analysis, statistical analysis, and spatial analysis, this study measures the extent of the exposure risk reduction of different income groups and different racial and ethnic groups. Using three indicators of exposure risk, the study incorporates two distinct and complimentary datasets at a granular level to capture insights which otherwise would have been overrun by coarser scale analysis. The first indicator captures the number of trips based on a ZIP code-to-ZIP code origin-destination (O-D) network analysis. This indicator provides insights regarding cross-ZIP codes transmission risk of the virus. The greater the inflow measure of the number of trips to nodes within ZIP codes, referred in this paper as the in-degree flow of a ZIP code, the higher the exposure risk of residents in that ZIP code to virus transmission from other ZIP codes. The second indicator examines the points of interest (POIs)-to-census block group (CBG) networks to capture the exposure risk of contact at POIs. The third indicator captures the exposure risk from previous transmission at POIs back to their home CBGs. These three indicators provide distinct measures as proxies for evaluating exposure risk reductions afforded by NPIs and enable us to examine the disparities among vulnerable sub-populations. The spatiotemporal context of the study comprises of five US locations: (1) Cook County (Chicago), Illinois, (2) Harris County (Houston), Texas, (3) New York City, New York, (4) Los Angeles County (Los Angeles), California, (5) King County (Seattle), Washington, recorded between January 1, 2020 through July 31, 2020.

**Methods**

**Description of Mobility Data and Population Activity**
The research uses two distinct and complimentary datasets to understand mobility patterns and population activity of communities as they relate to exposure risk to COVID-19 (Figure 1). The inflow measures of StreetLight Data and the POI visits of SafeGraph were examined through network, statistical analysis, and spatial analysis (Figure 2). The mobility data was obtained from StreetLight Data, which "harnesses smartphones as sensors to measure vehicle, transit, bike, and foot traffic." Per month, the company processes and aggregates approximately 40 billion anonymized records which include more than 5 million miles of roadway, sidewalk, and bike lanes. In particular, this research study aggregated the mobility data to a ZIP code-to ZIP-code origin-destination network model to analyze the number of inflow trips across ZIP-codes. As such, the first exposure risk indicator is a measure of increased transmission across ZIP codes through the inflow measure (in-degree values) of ZIP codes. The greater the in-degree flow of a ZIP code, the higher the exposure risk of residents in that ZIP code to virus transmission from other ZIP codes.

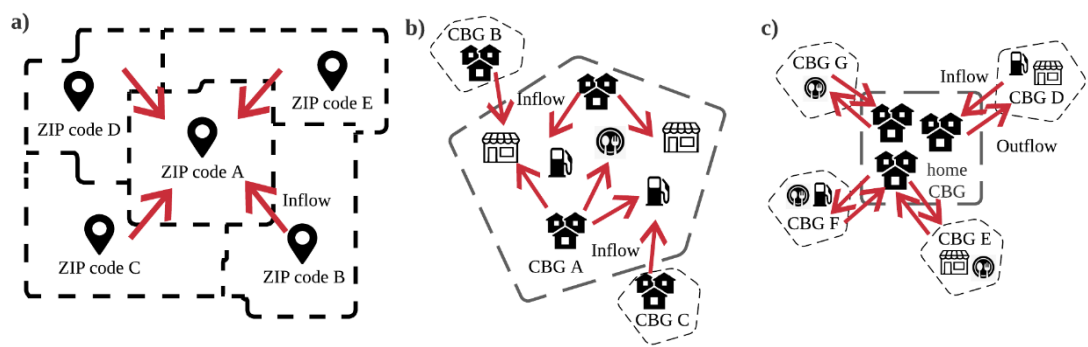

Figure 1. Mobility patterns and population activity yield exposure risk indicators. Exposure risk indicators are measured through a) inflow measures from ZIP codes B, C,D, E to ZIP code A, b) number of inflow visits of contacts at POIs in CBG A with more inflow visits from CBG B and CBG C to POIs in CBG A, c) previous transmission at POIs in CBG D,E, F, G to home CBG

In addition, the population activity is examined based on the SafeGraph data, which is "the most accurate points of interest (POIs) and store location geofences for the United States," and contains more than 8.2 million high-precision POIs. Using this dataset, we created a census block group to point-of-interest network model to examine the number of visits to POIs as well as the home CBG of visits. The second exposure risk indicator is the extent of contact at POIs measured as the number of visits to POIs in different CBGs. The greater the number of visits to the POIs of a particular CBG, the higher the exposure risk of residents in that CBG. The third exposure risk is the previous contact made at POIs and transmission to home CBGs measured as the number of visits from different home CBGs. The greater the number of visits made to various POIs by residents of a particular CBG, the higher the exposure risk of the residents of the CBG.

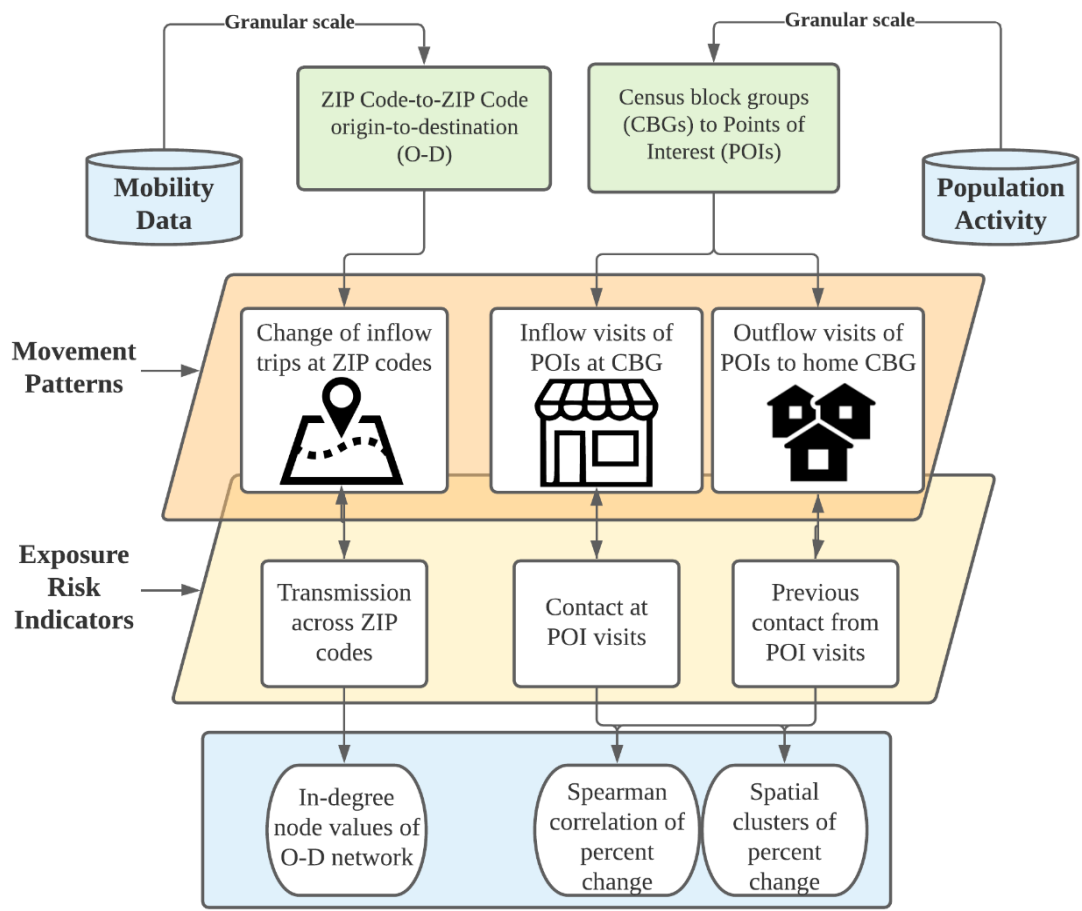

**Figure 2.** Methodological Framework for mobility and population activity

Five US counties with large cities were selected as the study areas: (1) Cook County (Chicago), Illinois, (2) Harris County (Houston), Texas, (3) New York City (comprising five counties) (4) Los Angeles County (Los Angeles), California, (5) King County (Seattle), Washington. The first four locations are the four most populated cities in the United States; Seattle was also included because it was the city with the first recorded instance of an individual diagnosed with the COVID-19 virus. The selected locations are also widespread across regions of the United States to contrast differences in impact.

The analysis time period was January 1, 2020 through July 31, 2020, which was generally the first wave of the pandemic, whereas the majority of NPIs such as shelter-in-place orders took place during mid-March (as shown in the Supplemental Information Sub-Section A) and were mostly released by late May and early June 2020. To normalize the data, a baseline level for mobility movement and average POI unique visits was established between January 13, 2020 and February 4, 2020. The baseline was selected after the early weeks of the year to give a more stable comparison that was not biased towards the heavy tourism and movement of the typical holiday period. After filtering, StreetLight Data had 83,460,324 total datapoints and SafeGraph had 354,034 total data points for the five locations. For StreetLight Data, Chicago had 11,638,220 points; Houston had 29,789,279 points; New York City had 15,485,617 points; Los Angeles had 22,712,811 points; and Seattle had 3,834,397. For the SafeGraph, Chicago had 55,292 points; Houston had 52,776 points; New York City had 72,304 points; Los Angeles had 146,037 points; and Seattle had 27,625 points.

Filtered data was merged with demographic information, specifically income and racial and ethnic groups, at a granular level using American Community Survey (ACS) [26, 27]. For both the mobility data and population activity data, income groups were divided into six levels of median income groups: (1) < $20,000, (2) $20,000–$49,999, (3) $50,000–$99,999, (4) $100,000–$149,999, (5) $150,000–$199,999, and (6) ≥$200,000 [27]. In addition, the mobility dataset examines six different racial and ethnic groups. These designations are established by ACS data, which follows the United States Office of Management and Business (OMB) standards: (1) White-only, (2) Black or African American, (3) American Indian (also known as Native American) or Alaska Native, (4) Asian, (5) Native Hawaiian/ Other Pacific Islander, and (6) Hispanic or Latino. The first five racial and ethnic populations total to 100 percent while the Hispanic or Latino populations overlap with the remaining racial and ethnic populations. The population activity dataset examined the percentage of (1) White-only and (2) non-White households which total to 100 percent [26] Median values of the sociodemographic information for each urban location can be found in Table A1 of the Supplemental Information.

**Analysis of ZIP code-level mobility network**
The mobility data describes the hourly number of trips for each pair of O-D links among across all ZIP-code areas. From the hourly origin-destination travel data, the hourly O-D network can be constructed, with the centroid points of all ZIP-code areas considered as nodes. At time $t$, if there exist trips from ZIP code $i$ to ZIP code $j$, a link between these two points will be constructed, and the number of trips, $N_{i,j}(t)$, is assigned as the weight of this link. For each node $i$, the weighted degree $D_i$ can be calculated by summing all the weights of links connected to node $i$ using Equation 1. This allowed us to calculate the movement inflows using n-degree and outflows using out-degree measures for each ZIP code. The variation in the in-degree measure could indicate the extent to which residents of a particular ZIP code could be at risk of transmission of the virus from other ZIP codes. Hence, we examined the in-degree measure across different ZIP codes and their variation due to NPIs compare income groups and different racial and ethnic groups.

$$D_i(t) = \sum_j N_{i,j}(t) \tag{1}$$

where, $D_i(t)$ is the in-degree/out-degree, and $N_{i,j}(t)$ is the number of trips of starting node ($i$) and ending node ($j$)

**Analysis of Population Activity Fluctuations**

SafeGraph collects data on the number of unique visits to each POI, (i.e., grocery stores, restaurants, education centers, and medical facilities), based on anonymized cell-phone data. The percent change of visits from the baseline of visits was calculated as shown in Equation 2.

$$PC_i = \frac{Visits_i - Baseline}{Baseline} * 100\% \tag{2}$$

where, $PC_i$ is the percent change of visits, $Visits_i$ is the number of visits at week i, and Baseline is average visits between January 13, 2020 and February 3, 2020. The variation in the total number of visits to POIs in a particular CBG could indicate the level of contact (and potential transmissions) at POIs. Hence, we examined the variation in visits to POIs in each CBG as a risk reduction indicator due to NPIs. Then, as the income and racial and ethnic data were grouped into categories, the Spearman correlation was used to determine disparity of socially vulnerable populations. For each individual week, the correlation between the percent change of visits to the baseline and the different social groups were calculated. For example, it determined whether POI visits at lower-income CBGs had lower percent change and more exposure risk when compared to POI visits at higher-income CBGs at statistically significant levels ($p<0.05$).

**POI-CBG network analysis**

In addition to measuring the population activity fluctuations at POIs, when possible, SafeGraph provides the number of visits from the home CBG of the visitor. Home CBGs were limited to those within the county; thus, captured visits originating in a different county were not used in the analysis. This kept the exposure risk indicator as a measure within the residents of the county. Thus, a network analysis can be created from the link between home CBG to POI. Similar to the calculation of population activity analysis, the percent change of mobility was calculated (Equation 2) along with Spearman correlation. This indicates which home CBGs had greater transmission from previous POI visits and greater risk to bringing exposure risk to their home CBG by measuring the number of POI visits. In this case, the Spearman correlation determined whether residents from lower-income CBGs had a lower percent change of visiting different POIs in the community when compared to residents from higher-income CBGs at statistically significant levels ($p <0.05$).

**Spatial Clustering of Exposure Risk Indicators**

In addition, bivariate Moran's I statistic was calculated to examine the spatial autocorrelation, or spatial clustering, of the exposure risk indicators. In particular, we are interested in understanding if there were areas of extreme exposure risk for contact at POIs and transmission to home CBGs to detect hot-spots or cold-spots of population activity and movements. Such additional insights could reveal areas of high vulnerability and low vulnerability within urban communities through a spatial demonstration, which is otherwise not shown through correlations. Clusters are generated as shown in Equation 3 using two variables— (1) the percent change to the baseline and (2) type of social groups—and are recorded when p-values are statistically significant (p<0.05). Clusters can either be high-high (H-H), high-low (H-L), low-high (L-H), or low-low (L-L). H-H clusters represent areas of high socially vulnerable populations (low income or racial and ethnic minority groups) and high exposure risk (greater mobility) and represent hot-spots in the community. The L-L clusters represent areas of low socially vulnerable populations (high-income or White-only populations) and low exposure risk (less mobility) and represent cold-spots in the community.

$$I_t = \frac{R * \sum_{i=1}^{R} \sum_{j=1}^{R} w_{ij} x_{i*} x_j}{R_b \sum_{i=1}^{R} x_i^2} \quad (3)$$

where, $I_t$ is the Moran's I statistic, $R$ represents the number of regions (CBGs) in the dataset; $w$ is the weight of the socially vulnerable population (income group or racial and ethnic group); $x$ is the percent change at CBG (*i*) and CBG (*j*)

**Results**

**Transmission across ZIP codes for O-D Network**

The first exposure risk indicator accounts for the transmission across ZIP codes based on analysis of the O-D network derived from the mobility dataset. An example of the O-D Network, which shows the percent change from inflow measures and outflow measures from the established baseline, can be found in Figure A1 of the Supplemental Information. In addition, Figures 3 and 4 show the in-degree results, or the inflow measure from ZIP code to ZIP code, for different income groups and different racial and ethnic groups. In-degree values were normalized based on the volume of trips divided by the baseline number of trips for each social group to account for uneven distribution of the number of trips for each socio-demographic group. Generally, in-degree values returned to baseline levels (equal to or greater than 1.0) by June–July.

Following the implementation of NPI between March 19 and March 23, 2020, there is a notable divergence of in-degree values among the different social groups. Thus, ZIP codes with higher income groups had greater in-degree variations, while ZIP codes with lower income groups had less changes in-degree variations. Generally, and for all socio-demographic groups, in-degree values dropped after March 16, 2020, but returned to baseline values for all urban cities by the end of July. However, the initial drop of in-degree values was less for lower-income residents traveling to different ZIP codes, indicating a lower exposure risk reduction due to the NPIs for those populations.

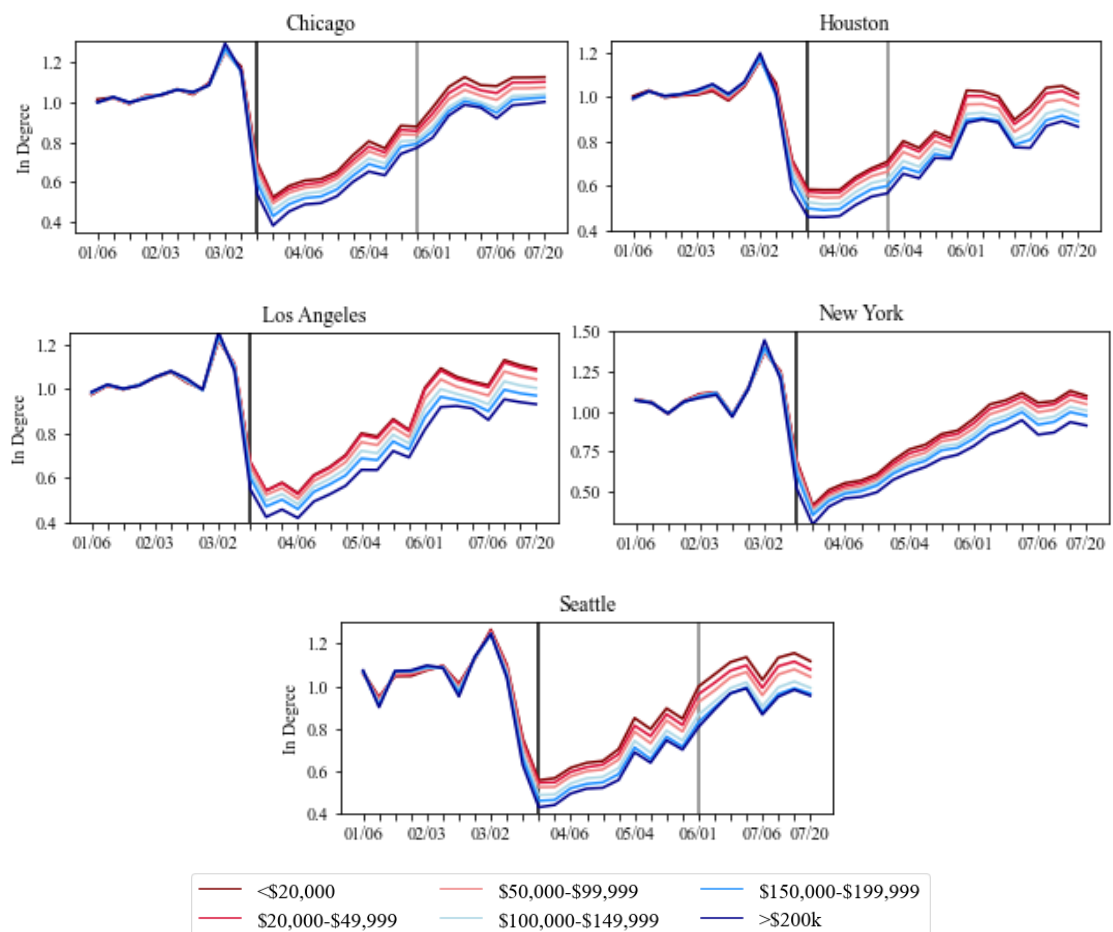

**Figure 3.** Variation in inflow of trips in ZIP codes for different income groups. In-degree values are normalized for each socio-demographic group to account for uneven distribution. The vertical black line indicates the week that an NPI shelter-in-place order was implemented in each county, and, if applicable, the vertical gray line indicates the week that the NPI shelter-in-place order was lifted.

Regarding the racial and ethnic groups, there is a notable divergence of in-degree values either before or during the implementation of NPIs. Across five urban locations, White-only populations had the greatest drop of in-degree values, meaning these populations had, comparatively, one of the lowest exposure risks (i.e., greatest reduction in trips to the ZIP codes from other ZIP codes). Native Hawaiian/ Other Pacific Islander and American Indian or Alaska Native populations showed virtually no change in their in-degree values, and thus, the greatest exposure risk comparatively for traveling cross ZIP codes. By comparison, Black or African American and Asian populations had fewer in-degree variations than White-only populations, but greater in-degree variations than Native Hawaiian/ Other Pacific Islander and American Indian or Alaska Native populations. To clarify, the percentage of racial and ethnic groups for populations of White-only, Black or African American, American Indian or Alaska Native, Asian, and Native Hawaiian/ Other Pacific Islander total to 100 percent; however, the percentage represented by the Hispanic ethnicity population overlaps with other categories. Results for Hispanic populations were mixed. While the group initially had the third lowest decrease of in-degree values, this initial drop stayed consistent throughout the timeframe to conclude with the greatest drop of inflow measures at the end of the analysis period. This indicates that although the Hispanic population initially had a high exposure risk, or low reduction in trips to ZIP codes, exposure risk decreased over time.

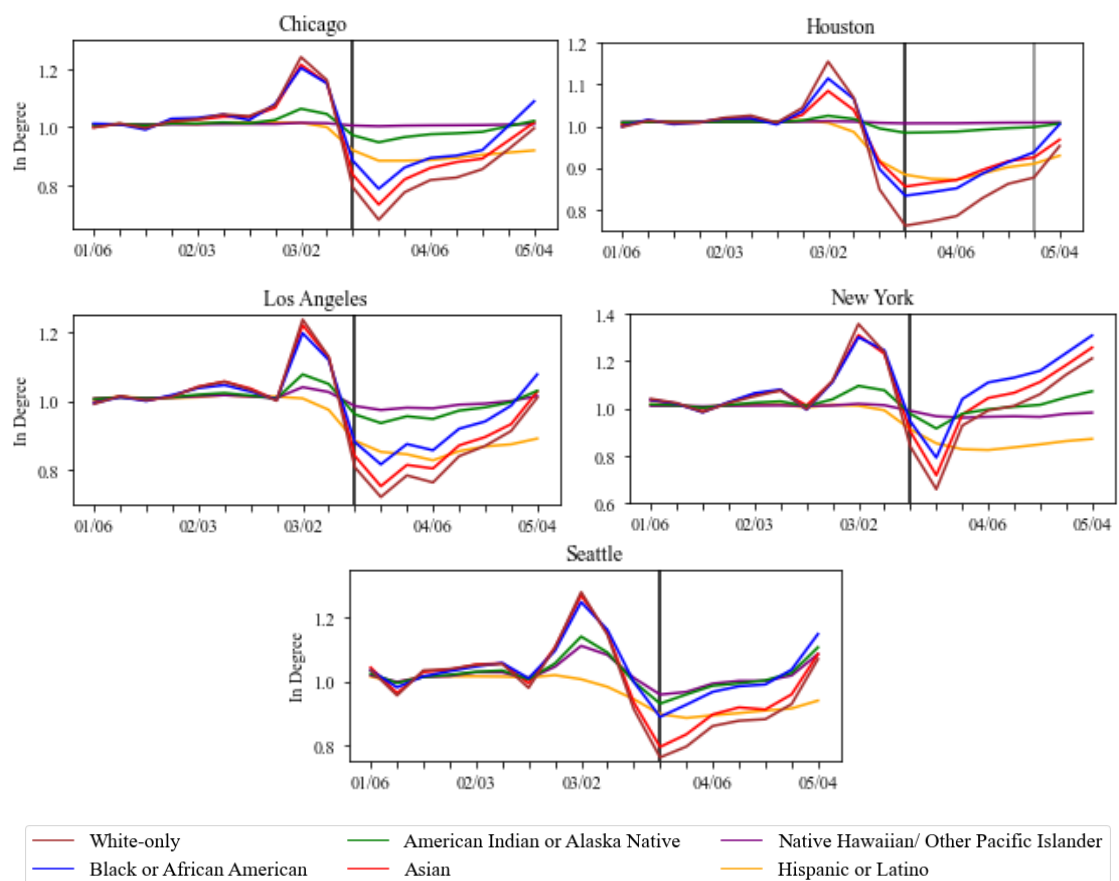

**Figure 4.** Variation in inflow of trips in ZIP codes for different racial groups. In-degree values are normalized for each socio-demographic group to account for uneven distribution. The percentage of each racial and ethnic group, except Hispanic, totals to 100% population. Hispanic ethnicity overlaps with the other racial and ethnic groups. The vertical black line indicates the week that an NPI shelter-in-place order was implemented in each county, and, if applicable, the vertical gray line indicates the week that the NPI shelter-in-place order was lifted.

The second exposure risk indicator accounts for the possible contact at all POIs in a particular CBG based on the percent change of POI visits from the baseline level (Figure 5) derived from the population activity dataset. The higher the number of POI visits, the greater the exposure risk. Generally, the POI percent change did not return to baseline levels but rather stayed below -40% from the baseline, which differs from the mobility dataset. Though mobility within a community, or number of trips within the community, may have returned to a steady state, this does not mean that people are physically entering the locations, as some businesses and organizations offered curbside pickup, delivery, and virtual services. As shown in Figure 5, following the implementation of NPIs, there was a notable difference in the visits to POIs in lower-income CBGs when compared to the visits to POIs in higher-income CBGs, where lower income areas had less percent change from the baseline due to NPIs compared to higher-income areas. This result indicates a lower exposure risk reduction in lower-income CBGs due to NPIs compared to that of higher-income CBGs. The release of NPIs in Chicago, Houston, and Seattle was associated with different results in the urban locations. After the lifting of the stay-at-home policy in Chicago and Seattle, CBGs with a median income >$200,000 had less percent change of POI visits from the baseline or less reduction in risk exposure. The release of NPI, however, did not dramatically influence the percent change of POI visits in CBGs of Houston, meaning lower-income CBGs had more modest decrease in reduction exposure compared to higher-income CBGs.

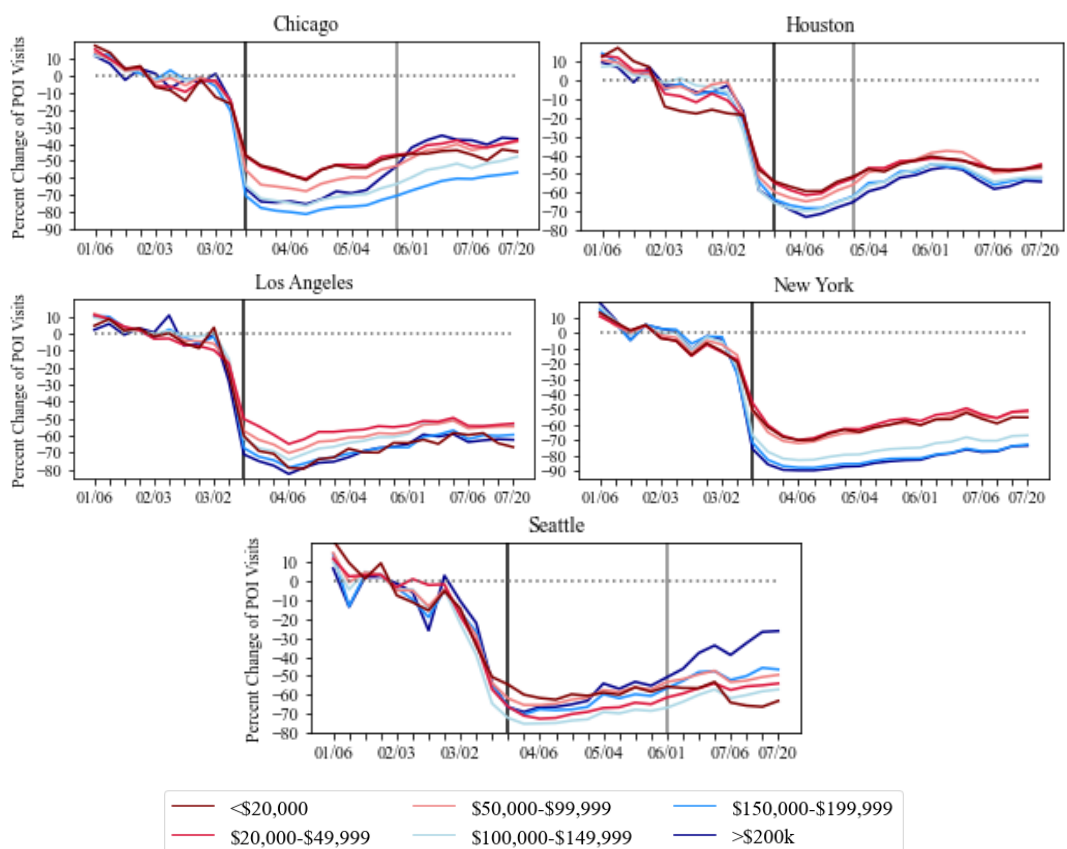

**Figure 5.** Percent change in POIs visits in CBGs for different income groups. Percent change values are normalized for each socio-demographic with an established baseline. The vertical black line indicates the week that a NPI shelter-in-place order was implemented in each county, and, if applicable, the vertical gray line indicates the week that NPI shelter-in-place order was lifted.

We also performed Spearman correlation analysis by calculating the correlation coefficients and p-values, where $p<0.05$ is statistically significant, for the percent change of POI visits in different CBGs and different socio-demographic populations:

median income groups and the percentage of White-only and non-White populations (see Figure A2 in Supplemental Information). Positive correlation values signify that CBGs with greater income levels or greater percentage of White populations had lower percent change of POI visits, while negative correlation values signify these CBGs had higher percentage change of POI visits. Prior to the implementation of NPIs, Chicago, New York City, Houston, and Los Angeles generally had positive correlations, meaning that POIs in CBGs with higher median income and higher percentage of White populations received more visits. In particular, Chicago and New York City had positive correlation values (between 0.10 and 0.30) and statistically significant p-values for median income and percent change of POI visits, which further supports higher-income CBGs were associated with more POI visits before the implementation of NPIs. This would have indicated higher-income CBGs had greater risk exposure to contact at POIs. After the implementation of NPIs, however, these correlation values flipped from positive to negative, which indicates a major shift in the fluctuations of population activity at the urban locations. Between March 16 and June 1, 2020, Chicago, New York City, and Houston had negative correlation values (between -0.15 and -0.30) at statistically significant p-values, and Los Angeles had negative correlations (approximately -0.05) at statistically significant p-values for the analysis of income groups and percent change of POI visits. After June 1, 2020, only New York City retained negative correlation values at statistically significant p-values. Additionally, New York City, Chicago, and Los Angeles had low negative correlation values (between -0.05 and -0.25) for the analysis of the percentage of White-only populations and percent change of POI visits; however, only Chicago and New York City had these at statistically significant p-values. The correlations indicate an association that CBGs of lower median income in Chicago, New York City, Houston, and Los Angeles had less reduction in exposure risk only after NPIs, while CBGs of greater percentage of minority populations in Chicago and New York City had less reduction in exposure risk after NPIs.

**Previous transmission from POIs to home CBGs**
The third exposure risk indicator accounts for previous transmission from POIs to home CBGs (Figure 6). The greater the number of visits made to various POIs by residents of a particular CBG, the higher the exposure risk of the residents of the home CBG. The measure used for this indicator is the total visits from a home CBG to all POIs in the POIs-to-CBG network. Like the results of contact at POIs analysis, percent change of home CBGs stayed below -40% from the baseline, and there were instances of exposure risk disparity across all urban locations due to NPIs. Residents from home CBGs with lower-income populations had lower percent change in their visits compared to their baseline than higher-income CBGs. This suggests that lower-income households were less able to reduce their exposure risk because they visited comparatively more POIs. After the release of their respective NPIs, the exposure risk disparity, or the difference in percent change of visits from lower-income CBGs and from higher-income CBGs, appeared to decrease for both Chicago and Seattle. The results generally show that residents from different CBGs were reducing their exposure risk at the same level, which suggests that residents from lower-income CBGs were approaching a similar exposure risk to that of higher-income CBGs. The percent change of total POI visits for different home CBGs in Los Angeles also appeared to converge despite not having a release of NPIs. In contrast, Houston still had a notable difference in the percent change of visits from lower-income CBGs when compared to higher-income CBGs well after the release of the NPI. New York City, which did not have a release of their NPIs, had the greatest difference of their percent change visits to POIs from lower-income CBGs and higher-income CBGs, and this difference actually increased by the end of the analysis period. This indicates that lower-income CBGs were visiting more total POIs, and thus, these areas were less able to reduce their exposure risk.

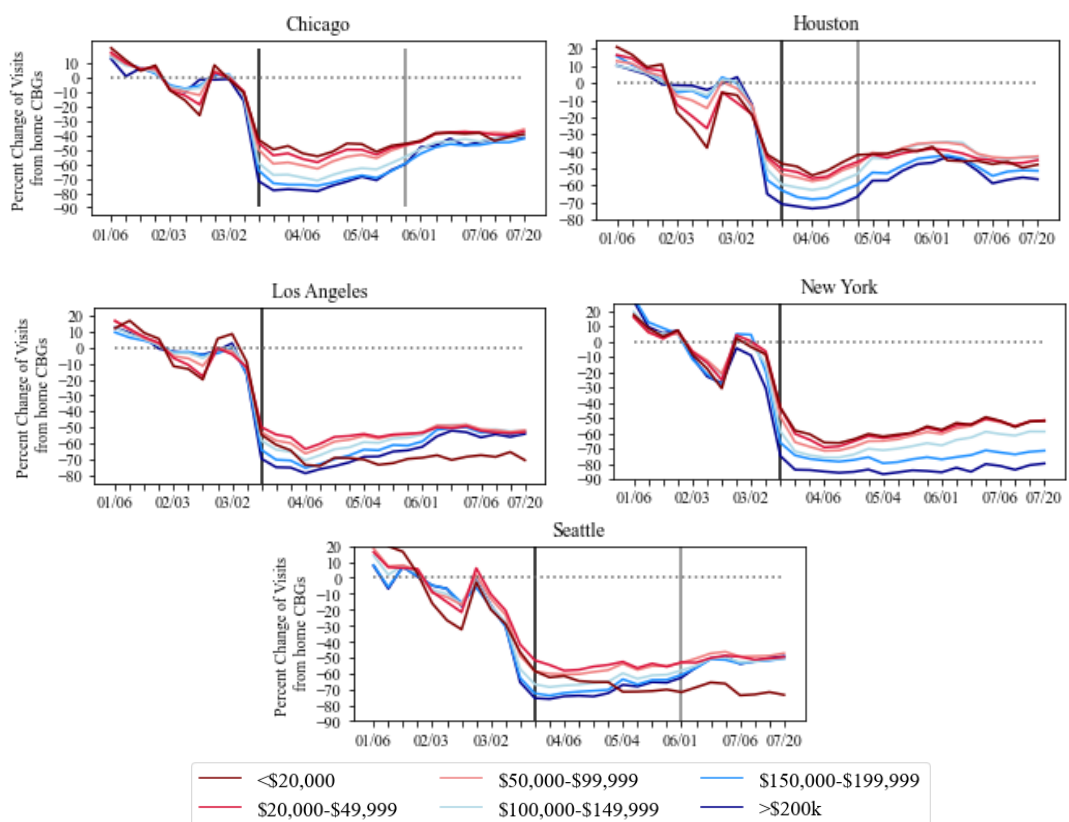

**Figure 6.** POI percent change of total visits of POIs to home CBGs for different income groups. Percent change is normalized for each socio-demographic with an established baseline. The vertical black line indicates the week that a NPI shelter-in-place order was implemented in each county, and, if applicable, the vertical gray line indicates the week that the NPI shelter-in-place order was lifted.

Spearman correlation further supports the exposure risk disparity between socially vulnerable populations and the percent change of total POI visits from home CBGs (see Figure A3 in Supplemental Information). A positive Spearman correlation indicates that residents from lower-income CBGs or those with a greater percentage of minority population were better able to reduce their exposure risk and were traveling to fewer POIs, while a negative Spearman correlation indicates that residents from lower-income CBGs or those with greater percentage of minority population were less able to reduce their exposure risk and were traveling to more POIs. Approximately after the implementation of NPIs in Chicago, Houston, and Los Angeles, there is a flip from positive to negative correlations at statistically significant p-values for different income groups (between -0.20 and -0.45) and non-White populations (between -0.05 and -0.45), which indicates a major shift in mobility patterns for the CBG-POI network. Even New York City and Seattle, which showed no notable distinctions in mobility between CBGs of different socio-demographic characteristics, had low negative correlations at statistically significant p-values after implementation of NPIs for different income groups (between -0.15 and -0.30) and non-White populations (between -0.20 and -0.45). Such correlation values remained for all the urban locations until the beginning of June 2020, with New York City still having the highest correlations of exposure risk disparity at statistically significant p-values. This indicates that residents of home CBGs with socio-demographic characteristics of lower median income and higher percentage of minority populations had a greater exposure risk due to increased total visits to POIs.

**Spatial Mapping of High Exposure Risk Areas**

In addition to performing Spearman correlation on the population activity and CBG-POI network, spatial mapping of the bivariate Moran's I statistic demonstrates the autocorrelation of the percent change of visits at POIs and total visits from home CBGs. As a visual evaluation, we compared a time period before the implementation (January 27 through February 2, 2020), after the implementation (April 6 through April 12 2020), and, if applicable, after the release of NPIs (June 15 through June 21) for different socially vulnerable populations. As previously stated, the H-H clusters represent areas of high socially vulnerable populations, or low income or racial and ethnic minority groups, and high exposure risk, or lower percent change of visits at POIs and total visits from CBGs. The L-L clusters represent areas of low socially vulnerable populations, high-income or White-only populations, and low exposure risk, or greater percent change of visits at POIs and total visits from CBGs. Such clusters were statistically significant at $p<0.05$. The paper presents an example of this spatial analysis by comparing the H-H clusters and L-L clusters of the urban locations regarding the percent change of total visits from home CBGs. It also compares significant spatial clusters of lower income and higher income groups regarding the percent change of total visits from home CBGs to illustrate potential areas in the community with greater exposure risk from traveling to more POIs, which otherwise would have not been visually seen with Spearman correlation analysis.

Significant spatial results, or p-values $< 0.05$, were found for transmission of previous contact at POIs to home CBGs in the CBG-POI network (Table 1) but not for the contact at POIs for population activity fluctuations (as shown in Table A2 of the Supplementary Information). Table 1 shows the Moran's I Statistic and the number of H-H clusters and L-L clusters for the CBG-POI network. After the implementation of NPIs, H-H and L-L clusters are increased for Chicago, Los Angeles, New York City, and Seattle, but not for Houston. In these urban locations, there are statistically significant clusters, or p-values $< 0.05$, of home CBGs with lower median income and lower percentage of White-only populations which were visiting more POIs in comparison to CBGs with higher median income and lower percentage of White-only populations. This indicates there may be a spatial component to the increased exposure risk of home CBGs. Figure 7 visually demonstrates the shift in the number of statistically significant H-H and L-L clusters (p<0.05) for Chicago (left) and New York City (right) for January 27 through February 2, April 6 through April 12, and June 15 through June 21. In the case of Chicago, the number of H-H clusters and L-L clusters increased after implementation of NPI but decreased after the release of NPI. New York City, however, did not have a release of NPIs during the study period. The number of H-H clusters and L-L clusters increased after the implementation of NPIs and stayed consistent in the number of clusters throughout the study period.

**Table 1.** Bivariate Moran's I Statistic for Socially Vulnerable Populations for POIs to home CBGs

| City | Jan 27 to Feb. 2 | | | Apr. 6 to Apr. 12 | | | Jun. 15 to Jun. 21 | | |
|---|---|---|---|---|---|---|---|---|---|
| | Moran's I | H-H | L-L | Moran's I | H-H | L-L | Moran's I | H-H | L-L |
| **Income Groups** | | | | | | | | | |
| Chicago | .086 | 124 | 154 | .320 | 350 | 545 | .110 | 146 | 423 |
| Houston | .211 | 148 | 151 | .271 | 180 | 233 | .025 | 55 | 184 |
| Los Angeles | .046 | 118 | 248 | .219 | 305 | 720 | -0.005 | 124 | 475 |
| New York City | .039 | 128 | 215 | .223 | 278 | 616 | .257 | 223 | 771 |
| Seattle | .074 | 48 | 41 | .261 | 171 | 172 | .084 | 141 | 87 |
| **White and Non-White Groups** | | | | | | | | | |
| Chicago | .126 | 150 | 156 | .341 | 445 | 547 | .125 | 186 | 428 |
| Houston | .141 | 116 | 124 | .077 | 93 | 199 | -.087 | 59 | 169 |
| Los Angeles | .018 | 179 | 189 | .118 | 385 | 619 | -.020 | 399 | 267 |
| New York City | .033 | 159 | 178 | .225 | 422 | 584 | .178 | 178 | 89 |
| Seattle | .021 | 27 | 45 | .156 | 160 | 170 | .089 | 138 | 160 |

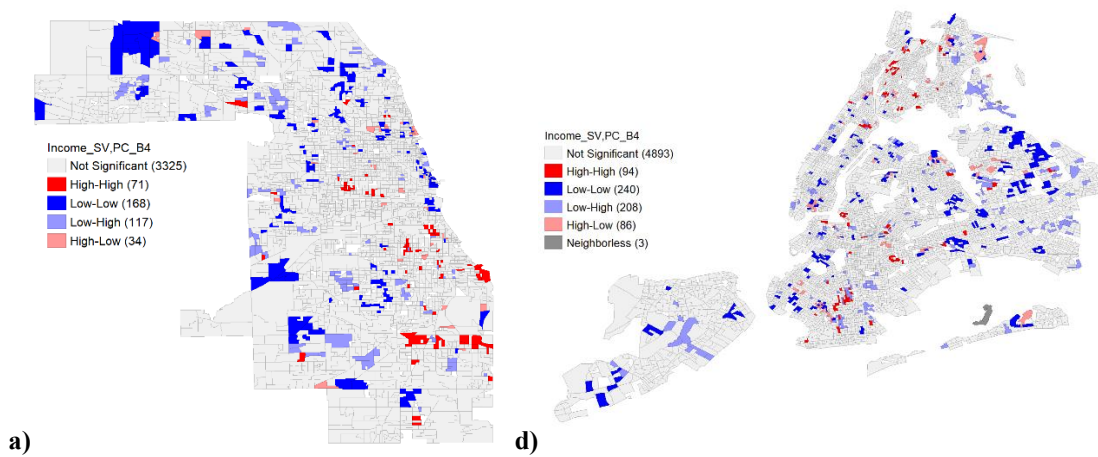

a)              d)

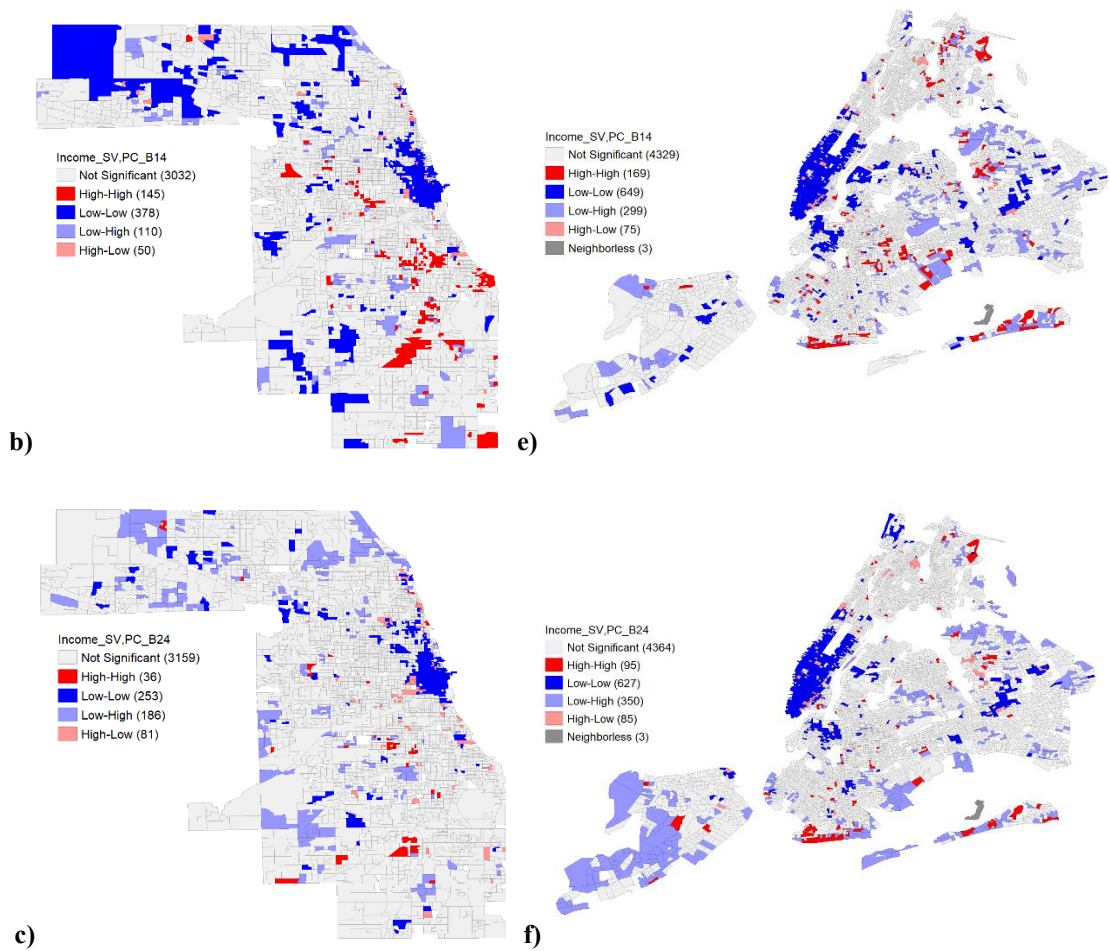

**Figure 7.** Visualization of exposure risk from contacts made in POIs transmitted to home CBGs. Spatial clusters are statistically significant for p<0.05. H-H clusters are shown in red; L-L are shown in blue. This shows the changing of significant clusters from Chicago (left) a) January 27 through February 2, 2020; b) April 6 through April 12, 2020, and c) June 15 through June 21; and New York City (right) d) January 27 through February 2, 2020; e) April 6 through April 12, 2020; and f) June 15 through June 21, 2020.

## Discussion

The resolution and nature of the datasets means that the results do not account for whether people were following all the safety guidelines of CDC, such as maintaining at the recommended six-foot distances and wearing approved masks, but the three exposure risk indicators can be used as proxies for measuring the protective actions of limiting movements around the community and around physical locations. It is also important to note, as with the majority of studies using mobility and location-intelligence, the data imbalance towards individuals and demographics owning smartphones. Given the quantity distribution of the mobility dataset and population activity dataset, the researchers feel that an accurate demographic was captured to measure the different patterns and behaviors of the five urban communities.

The COVID-19 pandemic has exacerbated systematic inequalities embedded in the health care system including poor access to medical services, costly medical treatments, inattention to underlying medical conditions, and misinformation and misunderstanding of safety policies [28-30]. The ever-growing body of research literature continues to uncover risk disparities associated with the pandemic, specifically the ability for different subpopulations to follow protective actions which reduce mobility around the community and limit exposure to the virus. Several studies have investigated different mobility metrics, such as the ability to stay at home [31], the intensity and duration of social distancing [32], exposure density of different neighborhoods [33], and the spatiotemporal contact density in particular industries [34], all of which reveal insights on the increased risk for socially vulnerable populations. However, there remains a knowledge gap of the underlying mechanisms which contribute to such disparities as well as a lack of granular analysis across multiple cities using different indicators of exposure risk.

Thus, in this study, we examined anonymized mobility data and population activity data from two distinct and complimentary datasets, to measure three indicators of exposure risk indicators to the COVID-19 virus: (1) transmission across ZIP codes, (2) contact at POIs, and (3) previous contact at POIs transmitted back to home CBGs for five urban locations (a) Cook County (Chicago), Illinois, (b) Harris County (Houston), Texas, (c) New York City, (d) Los Angeles County (Los Angeles), California, (e) King County (Seattle), Washington, between January 1, 2020, through July 31, 2020. First, the mobility data was used to create a ZIP code-to-ZIP code origin-destination network to record the inflow measures, or number of trips, to different nodes. Second, the population activity data recorded the fluctuations to the number of POI visits to physical locations. Third, the population activity data was used to establish a CBG-POI network to record the number of total POI visits from different home CBGs. The research study acknowledges the importance of non-pharmacological interventions, such as stay-at-home policies, which are effective in reducing the contact and transmission of the COVID-19 virus [35, 36]. After the implementation of stay-at-home policies of the county, results indicated a decrease in inflow measures of the mobility data and drop of percent change of POI visits in the population activity and CBG-POI network. The findings also suggest, however, a potential between the implementations of stay-at-home practices and the disproportionate impacts on different socio-demographic populations within the community through increased risk exposure.

The frequency of inflow measures of the mobility dataset, percent change to POI visits of population activity fluctuations, and percent change of total visits to POIs from different home CBGs through a POI-CBG network all indicate notable separations of mobility and visits patterns. The three exposure risk indicators support that socially vulnerable populations had higher exposure risk after the implementation of NPIs. Lower-income residents and certain racial and ethnic groups, particularly American Indian or Alaska Native, Pacific Islander, Black, and Asian populations, were less able to reduce their exposure risk across ZIP codes. In addition, Spearman correlation statistical analysis showed that POI visits in CBGs of lower income and higher percentage of minority populations had greater exposure risk and that home CBGs of lower income and higher percentage of minority populations had greater risk exposure at statistically significant p-values. Such disparity, however, was mixed or lessened over time and did not depend on the release of NPIs. These results were most notable for New York City and Chicago, although all urban locations showed instances of exposure risk disparity. In particular, spatial analysis also visualized statistically significant clusters, or p-values < 0.05, of high exposure risk and low exposure risk between different socio-demographic characteristics. POIs in CBGs were not clustered with significant p-values which indicates that on a spatial perspective, there was no difference in exposure risk for contact at POIs. However, there were statistically significant clusters, or p-values < 0.05, of home CBGs, which indicate a spatial component to the exposure risk of socially vulnerable CBGs particularly for New York City, Chicago, and Los Angeles, which must be further explored to understand the underlying mechanisms of exposure risk disparity regarding mobility and visits patterns.

In the conversation of social health disparities surrounding the pandemic, Chowkwanyun and Reed [37] discuss the importance of gathering data and information to develop a "precise picture of how vulnerability is distributed" while also emphasizing the importance of "[contextualizing] such data with adequate analysis." Though the findings highlight certain individuals and areas with high exposure risk to the virus because of an inability to reduce mobility and population activities, it is critical that researchers, policymakers, and the general public avoid stereotypes and stigmatization associated with socially vulnerable populations, which could delay resources, hinder participation, and limit voices in the recovery process. It is the responsibility for research studies to contextualize the possible factors influencing mobility disparity and exposure risk. While the results do capture and bring awareness to the vulnerability of different subpopulations, they also encapsulate the additional social disparities exacerbated by stay-at-home policies. Such policies, as previously implemented, do not consider that low-income groups and racial and ethnic minorities are more likely to work as essential and frontline workers in addition to having minimal pay, no sick leave, and being uninsured or underinsured [38]. Higher paying jobs may also be more flexible and accommodating to external shocks, such as the COVID-19 pandemic, and thus, they are able to offer work-from-home protocols [39]. On the other hand, those individuals with lower paying jobs would be more restricted in their work options, which could lead to many choosing between income and health.

Various studies have highlighted that socially vulnerable populations have been disproportionately impacted by the COVID-19 pandemic; however, the conversation of how to move forward from these significant impacts and, most importantly, prevent future ones must be centered on the notion that the ability to protect oneself is often a luxury perpetuated by external factors. Proper use of anonymized mobility data and population activity data can shed light on the effectiveness and equitability of closing and reopening policies. Although NPIs demonstrate to be effective in reducing mobility, there may be unintended consequences that must be addressed through careful governmental policies and protections which not only focus on direct connections to viruses but also the underlying mechanisms contributing to such exposure risk disparities.

**References**

1. Holshue, M.L., et al., *First Case of 2019 Novel Coronavirus in the United States.* New England Journal of Medicine, 2020. **382**(10): p. 929-936.
2. Abouk, R. and B. Heydari, *The Immediate Effect of COVID-19 Policies on Social-Distancing Behavior in the United States.* Public Health Reports, 2021. **136**(2): p. 245-252.
3. Badr, H.S., et al., *Association between mobility patterns and COVID-19 transmission in the USA: a mathematical modelling study.* Lancet Infectious Diseases, 2020. **20**(11): p. 1247-1254.
4. Nouvellet, P., et al., *Reduction in mobility and COVID-19 transmission.* Nat Commun, 2021. **12**(1): p. 1090.
5. Kraemer, M.U.G., et al., *The effect of human mobility and control measures on the COVID-19 epidemic in China.* Science, 2020. **368**(6490): p. 493-+.
6. Gozzi, N.T., Michele; Chinazzi, Matteo; Ferres, Leo; Vespignani, Alessandro; Perra, Nicola, *Estimating the effect of social inequalities on the mitigation of COVID-19 across communities in Santiago de Chile.* Nature Communications, 2021.
7. Gao, X.Y., et al., *Early Indicators of Human Activity During COVID-19 Period Using Digital Trace Data of Population Activities.* Frontiers in Built Environment, 2021. **6**.
8. Li, Q.C., et al., *Disparate patterns of movements and visits to points of interest located in urban hotspots across US metropolitan cities during COVID-19.* Royal Society Open Science, 2021. **8**(1).
9. Huang, X., et al., *The characteristics of multi-source mobility datasets and how they reveal the luxury nature of social distancing in the US during the COVID-19 pandemic.* International Journal of Digital Earth, 2021. **14**(4): p. 424-442.
10. Fraiberger SP, A.P., Candeago L, Chunet A, Jones NK, Khan MF, Lepri B, and L.L. Gracia NL, Massaro E, et al. , *Uncovering socioeconomic gaps in mobility reduction during the COVID-19 pandemic using location data.* 2020.
11. Fang, H.M., L. Wang, and Y. Yang, *Human mobility restrictions and the spread of the Novel Coronavirus (2019-nCoV) in China.* Journal of Public Economics, 2020. **191**.
12. Yabe, T., et al., *Non-compulsory measures sufficiently reduced human mobility in Tokyo during the COVID-19 epidemic.* Scientific Reports, 2020. **10**(1).
13. Wilson, B.S., *Overrun by averages: An empirical analysis into the consistency of social vulnerability components across multiple scales.* International Journal of Disaster Risk Reduction, 2019. **40**.

14. Coleman, N., A. Esmalian, and A. Mostafavi, *Equitable Resilience in Infrastructure Systems: Empirical Assessment of Disparities in Hardship Experiences of Vulnerable Populations during Service Disruptions.* Natural Hazards Review, 2020. **21**(4).
15. Dargin, J.S. and A. Mostafavi, *Human-centric infrastructure resilience: Uncovering well-being risk disparity due to infrastructure disruptions in disasters.* Plos One, 2020. **15**(6).
16. Finch, C., C.T. Emrich, and S.L. Cutter, *Disaster disparities and differential recovery in New Orleans.* Population and Environment, 2010. **31**(4): p. 179-202.
17. Mashoto, K.O., et al., *Socio-demographic disparity in oral health among the poor: a cross sectional study of early adolescents in Kilwa district, Tanzania.* Bmc Oral Health, 2010. **10**.
18. Szwarcwald, C.L., Souza-Júnior, P. R. B. D., Esteves, M. A. P., Damacena, G. N., & Viacava, F., *Socio-demographic determinants of self-rated health in Brazil.* Cadernos de Saúde Pública, 2005. **21**: p. S54-S64.
19. Thiele, T., et al., *Predicting students' academic performance based on school and socio-demographic characteristics.* Studies in Higher Education, 2016. **41**(8): p. 1424-1446.
20. Benitez, J.C., Charles; Yelowitz, Aaron, *Racial and Ethnic Disparities in COVID-19: Evidence from Six Large Cities.* Journal of Economics, Race, and Policy, 2020. **3**: p. 243-261.
21. Abedi, V., et al., *Racial, Economic, and Health Inequality and COVID-19 Infection in the United States.* Journal of Racial and Ethnic Health Disparities, 2021. **8**(3): p. 732-742.
22. Ossimetha, A., et al., *Socioeconomic Disparities in Community Mobility Reduction and COVID-19 Growth.* Mayo Clinic Proceedings, 2021. **96**(1): p. 78-85.
23. Borgonovi, F. and E. Andrieu, *Bowling together by bowling alone: Social capital and COVID-19.* Social Science & Medicine, 2020. **265**.
24. Sorg, L., et al., *Capturing the multifaceted phenomena of socioeconomic vulnerability.* Natural Hazards, 2018. **92**(1): p. 257-282.
25. Berrouet, L., C. Villegas-Palacio, and V. Botero, *A social vulnerability index to changes in ecosystem services provision at local scale: A methodological approach.* Environmental Science & Policy, 2019. **93**: p. 158-171.
26. Data, U.S.C., *American Community Survey, Table DP05*. 2019.
27. Bureau, U.S.C., *American Community Survey, Table S1903*. 2019.
28. van Dorn, A., R.E. Cooney, and M.L. Sabin, *COVID-19 exacerbating inequalities in the US.* Lancet, 2020. **395**(10232): p. 1243-1244.
29. Matrajt, L. and T. Leung, *Evaluating the Effectiveness of Social Distancing Interventions to Delay or Flatten the Epidemic Curve of Coronavirus Disease.* Emerging Infectious Diseases, 2020. **26**(8): p. 1740-1748.
30. Selden, T.M. and T.A. Berdahl, *COVID-19 And Racial/Ethnic Disparities In Health Risk, Employment, And Household Composition.* Health Affairs, 2020. **39**(9): p. 1624-1632.

31. Fu, X.Y. and W. Zhai, *Examining the spatial and temporal relationship between social vulnerability and stay-at-home behaviors in New York City during the COVID-19 pandemic.* Sustainable Cities and Society, 2021. **67**.
32. Garnier, R., et al., *Socioeconomic Disparities in Social Distancing During the COVID-19 Pandemic in the United States: Observational Study.* Journal of Medical Internet Research, 2021. **23**(1).
33. Hong, B., et al., *Exposure density and neighborhood disparities in COVID-19 infection risk.* Proceedings of the National Academy of Sciences of the United States of America, 2021. **118**(13).
34. Verma, R.Y., Takahiro; Ukkusuri, Satish V., *Spatiotemporal contact density explains the disparity of COVID-19 spread in urban neighborhoods.* Scientific Reports, 2021. **11**.
35. Tran, P., L. Tran, and L. Tran, *The Influence of Social Distancing on COVID-19 Mortality in US Counties: Cross-sectional Study.* Jmir Public Health and Surveillance, 2021. **7**(3).
36. Glaeser, E.G., Caitlin; Redding, Stephen J. , *How much does COVID-19 increase with mobility? Evidence from New York and Four Other U.S. Cities.* National Bureau of Economic Research, 2020.
37. Chowkwanyun, M. and A.L. Reed, *Racial Health Disparities and Covid-19-Caution and Context.* New England Journal of Medicine, 2020. **383**(3): p. 201-203.
38. Brown, E.A. and B.M. White, *Recognizing Privilege as a Social Determinant of Health During COVID-19.* Health Equity, 2020. **4**(1).
39. Ruiz-Euler, A.P., Filippo; Giuffrida, Danilo; , B. Lake, and I. Zara, *Mobility Patterns and Income Distribution in Times of Crises: US Urban Centers During the COVID-19 Pandemic.* SSRN, 2020.

**Acknowledgements**

The authors would like to acknowledge funding support from the National Science Foundation RAPID project #2026814: Urban Resilience to Health Emergencies: Revealing Latent Epidemic Spread Risks from Population Activity Fluctuations and Collective Sense-making, and Microsoft AI for Health COVID-19 Grant for cloud computing resources. The authors would also like to acknowledge that StreetLight Data and SafeGraph provided mobility and population activity data. Any opinions, findings, conclusions or recommendations expressed in this material are those of the authors and do not necessarily reflect the views of the National Science Foundation, Microsoft, SafeGraph or StreetLight Data, Inc.

**Data availability**

The data that support the findings of this study are available from SafeGraph and StreetLight Data, but restrictions apply to the availability of these data, which were used under license for the current study. The data can be accessed upon request submitted on StreetLight Data and SafeGraph. Other data we use in this study are all publicly available.

**Code availability**

The code that supports the findings of this study is available from the corresponding author upon request.

**Supplemental Information**

### A. Implementation of NPIs across five urban locations

Washington, King County
- Non-essential services closure 3/16/2020 ~ 5/5/2020
- Shelter-in-place 3/23/2020 ~ 6/1/2020

California, Los Angeles
- Non-essential services closure 3/19/2020 ~ 5/25/2020
- Shelter-in-place 3/19/2020 ~

Illinois, Cook County
- Non-essential service closure 3/21/2020 ~ 5/29/2020
- Shelter-in-place 3/21/2020 ~ 5/29/2020

New York City
- Non-essential services closure 3/22/2020 ~ 6/8/2020
- Shelter-in-place 3/22/2020 ~

Texas, Harris County
- Non-essential services closure 3/24/2020 ~ 5/1/2020
- Shelter-in-place 3/23/2020 ~5/1/2020

### B. Sociodemographic Information of Selected Urban Locations

Table A2. Sociodemographic Information from U.S Census Quick Tables

| Urban Locations | Median of the Household (in 2019 dollars) | White alone, percent | White alone, not Hispanic or Latino, percent |
|---|---|---|---|
| Chicago | $58,247 | 50.0% | 33.3% |
| Houston | $52,338 | 57.0% | 24.4 % |
| Los Angeles | $68,044 | 70.7% | 26.1% |
| New York | $63,998 | 42.7% | 32.1% |
| Seattle | $92,263 | 67.3% | 63.8% |

## C. In Degree/ Out Degree Percent Change Example

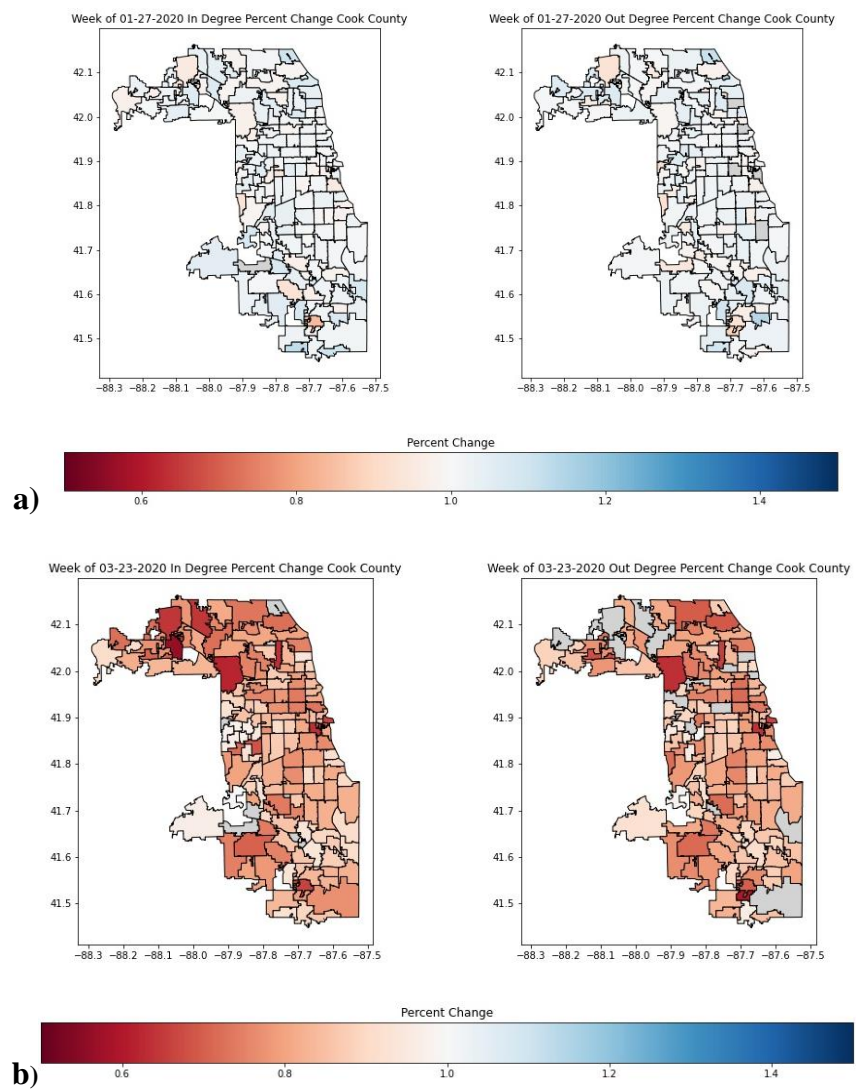

**Figure A1.** In Degree and Out Degree Percent Change of Cook County as an Example of O-D network. The images of the O-D network use the third week of Jan. (Jan. 20[th] – Jan 26[th]) as a baseline to the percent change of In Degree and Out Degree values, which are measures of inflow and outflow to the nodes. a) represents the week of Jan. 27[th] – Feb. 2[nd] while b) represents the week of Mar. 23[rd] – Mar. 29[th].

### D. Spearman Correlations for SafeGraph Data

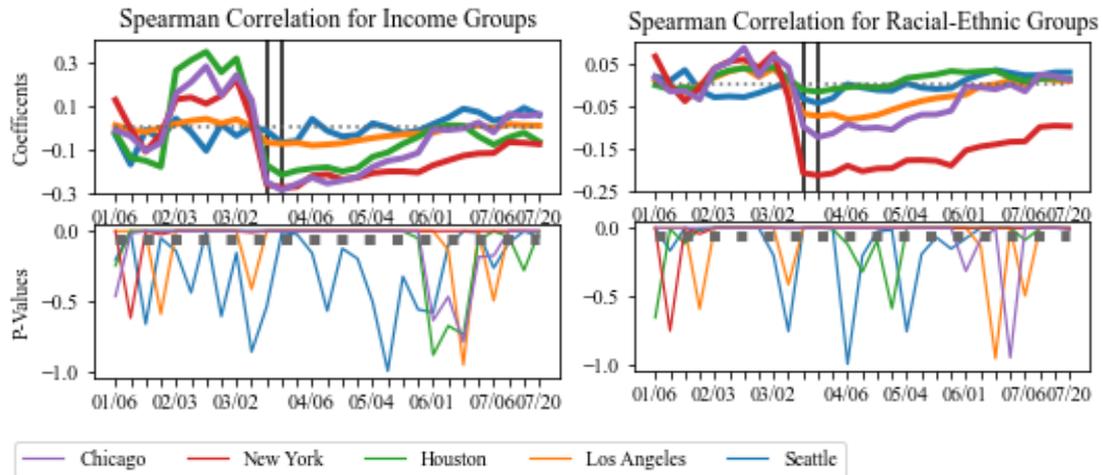

**Figure A2.** Spearman correlations of POIs visits at CBGs versus income and racial and ethnic attributes of different CBGs. The two vertical black lines indicate the weeks when NPIs were implemented since this varied for the five the urban locations. Income referred to median income of CBGs while racial and ethnic groups referred to percentage of non-White populations. P-values are statistically significant at p<0.05.

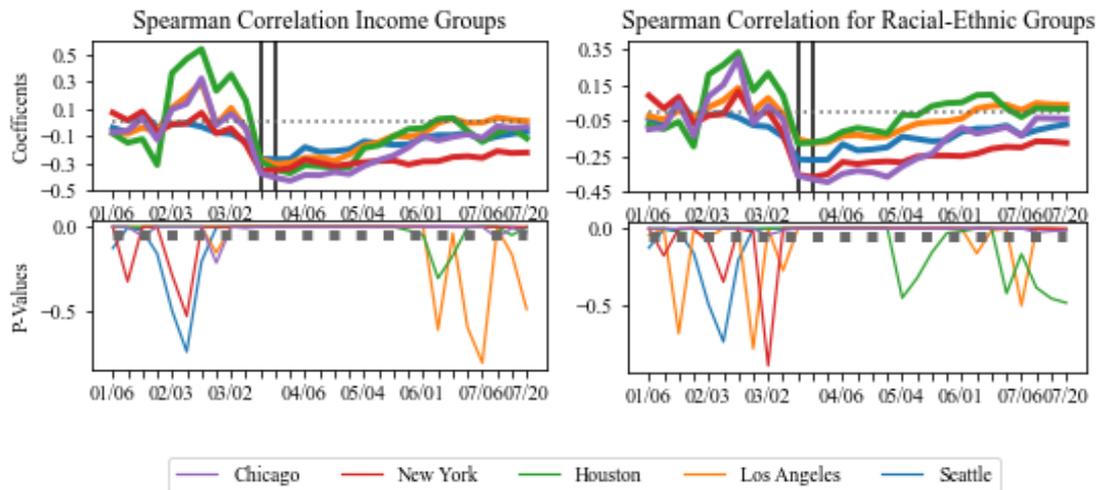

**Figure A3.** Spearman correlations of total visits to POIs in other CBGs versus the racial and income attribute of the home CBG. The two vertical black lines indicate the weeks when NPIs were implemented, since this varied for the five the urban locations. Income refers to median income of CBGs; racial and ethnic groups referred to percentage of non-White populations. P-values are statistically significant at p<0.05.

## E. Spatial Mapping for Contact at POIs

Table A2. Bivariate Moran's I Statistic for Socially Vulnerable Populations for Contact at POIs

| City | Jan. 27 to Feb. 2 | | | Apr. 6 to Apr. 12 | | | Jun. 15 to Jun. 21 | | |
|---|---|---|---|---|---|---|---|---|---|
| | Moran's I | H-H | L-L | Moran's I | H-H | L-L | Moran's I | H-H | L-L |
| **Income Groups** | | | | | | | | | |
| Chicago | 0.06 | 71 | 166 | 0.155 | 145 | 378 | -0.058 | 36 | 253 |
| Houston | 0.114 | 64 | 122 | 0.15 | 67 | 204 | 0.026 | 31 | 129 |
| Los Angeles | -0.016 | 75 | 254 | 0.09 | 156 | 545 | -0.04 | 23 | 308 |
| New York City | 0.02 | 94 | 240 | 0.223 | 169 | 649 | 0.13 | 95 | 627 |
| Seattle | 0.01 | 22 | 46 | -0.063 | 32 | 60 | -0.11 | 14 | 58 |
| **White and Non-White Groups** | | | | | | | | | |
| Chicago | 0.061 | 79 | 151 | 0.043 | 78 | 264 | -0.001 | 84 | 178 |
| Houston | 0.04 | 54 | 69 | -0.008 | 40 | 83 | -0.034 | 36 | 55 |
| Los Angeles | 0.022 | 176 | 202 | 0.032 | 183 | 258 | 0.006 | 171 | 190 |
| New York City | -0.012 | 126 | 159 | 0.063 | 168 | 418 | -0.014 | 152 | 414 |
| Seattle | 0.012 | 28 | 51 | -0.031 | 31 | 61 | -0.027 | 30 | 53 |